\documentclass[prd,a4paper]{revtex4}
\usepackage{epsfig}
\newcommand{\be}{\begin{equation}}
\newcommand{\ee}{\end{equation}}
\newcommand{\bea}{\begin{eqnarray}}
\newcommand{\eea}{\end{eqnarray}}
\newcommand{\bean}{\begin{eqnarray*}}
\newcommand{\eean}{\end{eqnarray*}}

\newcommand{\gapproxeq}{\lower
.7ex\hbox{$\;\stackrel{\textstyle >}{\sim}\;$}}
\newcommand{\lapproxeq}{\lower
.7ex\hbox{$\;\stackrel{\textstyle <}{\sim}\;$}}

\begin{document}

\bibliographystyle{unsrt}

\title{\bf On the near-threshold enhancement in $J/\psi\to \gamma X$ with $X\to\omega\phi$  }

\author{Qiang Zhao}

\affiliation{Institute of High Energy Physics, Chinese Academy of
Sciences, Beijing, 100049, P.R. China}

\affiliation{Department of Physics, University of Surrey,
Guildford, GU2 7XH, United Kingdom}

\author{Bing-Song Zou}

\affiliation{ CCAST (World Laboratory), P.O. Box 8730, Beijing
100080, P.R. China}

\affiliation{Institute of High Energy Physics, Chinese Academy of
Sciences, Beijing, 100049, P.R. China}

\date{\today}

\begin{abstract}

We investigate the possibility of producing the enhancement
observed in $J/\psi\to \gamma X$ with $X\to \omega\phi $ at BES by
intermediate meson rescatterings through $f_0(1710)\to PP\to
\omega\phi$, $f_0(1710)\to VV\to \omega\phi$, and $f_0(1710)\to
SS\to \omega\phi$. We find that intermediate meson rescatterings
can produce some enhancement near the $\omega\phi$ threshold.
Implications about the property of this enhancement are discussed.

\end{abstract}

\maketitle


\vspace{1cm}

\section{Introduction}

The observation of a near-threshold enhancement in $J/\psi\to
\gamma X$; $X\to \omega\phi$ at BESII~\cite{bes-2006} immediately
provokes discussions about its nature. This enhancement is
reported to favor $J^P=0^+$ in a partial wave analysis with a mass
and width of $M=1812\begin{array}{c} +19\\ -26
\end{array}(\mbox{stat})\pm 18 (\mbox{syst})$ MeV and
$\Gamma=105\pm 20(\mbox{stat})\pm 28(\mbox{syst})$ MeV,
respectively, and a production branching ratio, $B(J/\psi\to
\gamma X)\cdot B(X\to\omega\phi)=(2.61\pm 0.27(\mbox{stat})\pm
0.65(\mbox{syst}))\times 10^{-4}$.

The most interesting feature about $X(1810)$ is that it seems to
have a mass different from the previously observed $f_0$ states,
i.e. $f_0(1710)$ and/or $f_0(1790)$~\footnote{Signals for
$f_0(1790)$ are reported by BES Collaboration in
Ref.~\cite{bes-phi} in $J/\psi\to\phi f_0(1790)\to \phi K^+K^-$
and $\phi\pi^+\pi^-$. Its branching ratios to $\pi\pi$ are found
larger than to $K\bar{K}$ such that makes it distinct from
$f_0(1710)$. It has a width of about 270 MeV. Further experimental
confirmation is needed. }, and ``exclusively" couples to
$\omega\phi$ channel, while those known scalars are generally
below the $\omega\phi$ threshold. If one interprets the
enhancement as a Breit-Wigner resonance, it may lead to
implications of exotic meson
productions~\cite{amsler-close,close-tornqvist,horn-mandula} such
as glueball~\cite{bicudo,bugg-06}, hybrid~\cite{chao,hllz} and
four-quark state~\cite{liba}.

However, the sparse experimental information makes it difficult to
draw a decisive conclusion about its nature. A thorough
investigation of all possible interpretations of the data is thus
necessary, and mechanisms such as final state interactions should
also be inspected. The established $f_0(1710)$ at the affinity
makes it a candidate for such a consideration. PDG~\cite{pdg2004}
quotes $M=1714\pm 5$ MeV and $\Gamma_{tot}=140\pm 10$ MeV as the
average values of the mass and total width for the $f_0(1710)$. In
contrast, the recent data from BES~\cite{bes-gkk} with high
statistics indicate a relatively larger mass pole ($\sim 1740$
MeV) and total width ($\sim 166$ MeV). In the $f_0(1710)$ rest
frame, the production threshold for $\omega\phi$ will be at the
upper edge of the $f_0(1710)$ mass tail. It is known that
resonance off-shell effects can be significant due to the
energy-dependence of its decay width in a relativistic
Breit-Wigner form for a resonance propagating non-locally. For the
$f_0(1710)$ with the mass tail extended to about 1.8 GeV, the
possibility that the $f_0(1710)$ contributes via $J/\psi\to \gamma
f_0(1710)\to \gamma\omega\phi$ cannot be ruled out. Nonetheless,
the $f_0(1790)$~\cite{bes-phi}, if indeed a resonance, may also be
a candidate for producing the near-threshold enhancement in
$\omega\phi$ invariant mass spectrum.

The tantalizing feature of this enhancement is its presence in
$\omega\phi$ channel, which in general is OZI~\cite{ozi}
suppressed compared with other $VV$ decay channels such as
$\omega\omega$, $\phi\phi$, $K^*\bar{K^*}$, and $\rho\rho$. Strong
coupling to $\omega\phi$ simply implies the violation of the OZI
rule or a completely different production mechanism for
$X\to\omega\phi$. For the latter, proposals such as
Refs.~\cite{bicudo,bugg-06,chao,hllz,liba} are tackling some of
those points, for which a coherent picture is still absent. For
large OZI violations, the final state interactions due to
intermediate meson rescatterings seem likely to explain why $X$
only predominantly appears in $\omega\phi$, and is absent (at
least not strongly show up) in other channels such as
$K^*\bar{K^*}$, $\omega\omega$, $\rho\rho$ and $\phi\phi$.
Certainly, this consideration should be tested by numerical
studies, and our work in this paper is to provide some
quantitative evaluation of such a possibility.

The importance of the intermediate meson rescatterings (also
quoted as final state interactions) has been recognized in many
cases. Early studies by Lipkin show that the OZI-rule violations
can always proceed by a two-step process involving intermediate
virtual mesons~\cite{lipkin}. Detailed investigations by Geiger
and Isgur in a quark model highlight such a
correlation~\cite{isgur-geiger}. In particular, it is found that
systematic cancellations among the hadronic loops occur for the
$u\bar{u}\leftrightarrow s\bar{s}$ mixing in all nonets except
$0^{++}$, for which a general argument is given by Lipkin and
Zou~\cite{lipkin-zou}. The heavy quarkonium decays into light
hadrons provide a place for probing the intermediate meson
rescattering
mechanisms~\cite{locher-lu-zou,li-bugg-zou,cheng-chua-soni}. In
the charmonium energy region, the decay of $J/\psi\to \omega f_0$
and $\phi f_0$ have significant contributions from $J/\psi\to
K^*\bar{K}+c.c.\to\phi f_0$ and $J/\psi\to \rho\pi\to\omega f_0
$~\cite{zzm}. Similar phenomena are also observed in $\eta_c\to
\omega\phi$ as a major mechanism for the OZI-rule
violations~\cite{zhao-eta-c}.

For the $f_0(1710)$, its signal has been identified in different
processes~\cite{wa102,bes-phi,bes-omega}. It has large branching
ratios to $K\bar{K}$ from 0.38~\cite{pdg2004} to
0.6~\cite{bes-phi,bes-omega}. Taking into account that $\omega$
and $\phi$ both have sizeable couplings to $K\bar{K}$ and
$K^*\bar{K}+c.c.$, it is natural to conjecture that significant
contributions from the intermediate $K\bar{K}$ and/or
$K^*\bar{K^*}$ could be possible. Nevertheless, the scalar meson
exchange in $VV$ rescattering can also contribute, such as
$K^*\bar{K^*}\to\omega\phi$ via $\kappa$ exchange. Since
information about the $VVS$ interactions still lacks, we try to
gain some guidance about the couplings in the flavor SU(3)
symmetry. Correlated to this, it is then possible that
intermediate scalar meson rescatterings, such as
$\kappa\kappa\to\omega\phi$ via $K^*$ exchange, can contribute to
the cross sections. Although we do not have any experimental
information about the $f_0(1710)\to\kappa\kappa$ coupling, a test
of this channel should be useful for gaining some insights into
the $f_0(1710)$ properties. Similar to these considerations, the
$f_0(1790)$ observed in $\pi\pi$ channel will allow contributions
from $\pi\pi$ rescattering via $\rho$ exchange to $\omega\phi$.
Especially, note that $\phi\to \rho\pi + 3\pi$ has a sizeable
branching ratio~\cite{pdg2004}.

In brief, the intermediate meson rescattering prescription has
empirically accommodated as much as possible the available
information, from which we expect to learn more about the largely
unknown $X(1810)$. As follows, we will analyze three types of
rescatterings: i) intermediate pseudoscalar meson rescattering;
ii) intermediate vector meson rescattering; and iii) intermediate
scalar meson rescattering. We will study the lowest partial wave
contributions as leading processes. This should be sufficient for
gaining order-of-magnitude estimate at this stage. In Section II,
the formalism will be provided. We will discuss the numerical
results in Section III, and a brief summary is given in Section
IV.

\section{ Intermediate pseudoscalar meson rescattering}

\begin{figure}
\begin{center}
\epsfig{file=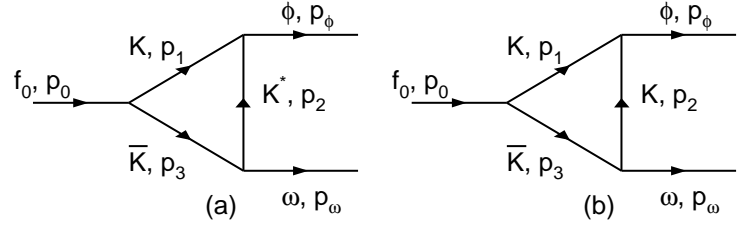, width=14cm,height=7.cm} \caption{Schematic
diagrams for intermediate meson rescatterings via (a) $K\bar{K}$
with $K^*$ exchange; and (b) $K\bar{K}$ with kaon exchange. }
\protect\label{fig-1}
\end{center}
\end{figure}

The intermediate pseudoscalar meson rescattering is illustrated by
Fig.~\ref{fig-1}, where (a) is for rescattering via vector meson
exchange, and (b), via pseudoscalar meson exchange.

\subsection{Rescattering via vector meson exchange}

The amplitude for the transition of Fig.~\ref{fig-1}(a) can be
expressed as
\be
{\cal M} = \int \frac{d^4 p_2}{(2\pi)^4}\delta^4(P_0-P_\phi
-P_\omega) \sum_{K^* pol.}\frac{T_a T_b T_c}{a_1 a_2 a_3} {\cal
F}(p_2^2) \ ,
\ee
with the vertex functions:
\bea
T_a &\equiv & i g_a M_0 ,
\nonumber\\
T_b &\equiv & \frac{i g_b}{M_\phi}\epsilon_{\mu\nu\xi\tau}
P_\phi^\mu \epsilon_\phi^\nu p_2^\xi \epsilon_2^\tau , \nonumber\\
T_c &\equiv & \frac{i
g_c}{M_\omega}\epsilon_{\lambda\iota\kappa\sigma}P_\omega^\lambda
\epsilon_\omega^\iota p_2^\kappa \epsilon_2^\sigma ,
\eea
where $g_a$, $g_b$ and $g_c$ are coupling constants at the meson
interaction vertices; Note that the tensor part of the vector
meson propagator will not contribute. The four-vectors, $P_0$,
$P_\phi$ and $P_\omega$, are momenta for the initial $f_0(1710)$
and final state $\phi$ and $\omega$ mesons, while $p_1$, $p_2$ and
$p_3$ are four momenta for the intermediate mesons, respectively.
Quantities, $a_1=p_1^2-m_1^2$, $a_2=p_2^2-m_2^2$ and
$a_3=p_3^2-m_3^2$, are the denominators of the propagators of
intermediate mesons.

By applying the Cutkosky rule, we have the decay amplitude
\be
\label{pp-rescat} {\cal M} = \frac{i g_a g_b g_c |{\bf p}_3|}{64
\pi^2 M_\phi M_\omega } (P_\phi\cdot
P_\omega\epsilon_\phi\cdot\epsilon_\omega
-P_\phi\cdot\epsilon_\omega P_\omega\cdot\epsilon_\phi) {\cal I} ,
\ee
with
\be\label{int-1}
{\cal I}\equiv\int d\Omega\frac{p_2^2  }{p_2^2-m_2^2}{\cal
F}(p_2^2)
 \ ,
\ee
where ${\cal F}(p_2^2)$ is the form factor introduced for the
off-shell vector meson ($K^*$ and $\rho$).

In Refs.~\cite{zzm,zhao-eta-c}, three cases have been studied for
the integration: (i) with no form factor; (ii) with a monopole
form factor, i.e. ${\cal
F}(p_2^2)=(\Lambda^2-m_2^2)/(\Lambda^2-p_2^2)$, where $\Lambda$ is
the cut-off energy; and (iii) with a dipole form factor, i.e.
${\cal F}(p_2^2)=[(\Lambda^2-m_2^2)/(\Lambda^2-p_2^2)]^2$.
Generally, the calculation results show large sensitivities to the
value of $\Lambda$. Therefore, to determine $\Lambda$ would
require sufficient experimental constraints. Unfortunately, such
constraints on $f_0\to VV$ are unavailable. We hence apply the
cut-off energy, $\Lambda=1.2\sim 1.8$ GeV, for the dipole form
factor calculations.

For the dipole form factor, the integration gives
\be\label{int-pp}
{\cal I}=2\pi \frac{A_s(\Lambda^2-m_2^2)^2}{B_s D_s^2}
\left\{\frac{2(A-D)}{(B-D)(1-D^2)}
-\frac{A-B}{(B-D)^2}\ln\frac{(1+B)(1-D)}{(1-B)(1+D)}  \right\}  \
,
\ee
where
\bea
A_s & \equiv & M_\phi^2+m_1^2-2E_\phi E_1 \ , \ \ A\equiv 2|{\bf
P}_\phi||{\bf p}_1| / A_s \ ; \nonumber\\
B_s & \equiv & A_s-m_2^2 \ , \ \ B\equiv 2|{\bf P}_\phi||{\bf
p}_1| /B_s \ ; \nonumber\\
D_s & \equiv & A_s-\Lambda^2 \ .
\eea

For $K\bar{K}$ rescattering, we adopt $B.R.(f_0(1710)\to
K\bar{K})=0.6$ to determine $g_a=(8\pi\Gamma_{f_0(1710)\to
K\bar{K}}/|{\bf p}_1|)^{1/2}$. Couplings $g_b=g_{\phi
K^{*+}\bar{K^-}}$ and $g_c=g_{\omega K^{*+}\bar{K^-}}$ are
determined in the SU(3) flavor symmetry limit:
$g_b=\sqrt{2}g_c=6.48$.

\subsection{Rescattering via pseudoscalar meson exchange}

In the pseudoscalar meson rescattering, it can also exchange
pseudoscalar meson such as $K$ and $\pi$ and then couple to the
final state $\omega$ and $\phi$ (see Fig.~\ref{fig-1}(b)). In this
process, considering that the couplings of $\phi K\bar{K}$ and
$\omega K\bar{K}$ both are large, it is necessary to investigate
the effects from this transition.

The $VPP$ coupling has a form of
\be
T_{VPP}=ig_{VPP} (p+p^\prime)\cdot \epsilon_v \ ,
\ee
where $p$ and $p^\prime$ are the four-vector momenta for the
incoming and outgoing pseudoscalars, respectively. We determine
the $g_{\phi K^+ K^-}$ via $\phi\to K^+ K^-$:
\be
g_{\phi K^+ K^-}=\frac{6\pi M_\phi^2}{|{\bf
p}_K|^3}\Gamma^{exp}_{\phi\to K^+K^-} \ ,
\ee
where $\Gamma^{exp}_{\phi\to K^+K^-}=2.09$ MeV is given by the
PDG~\cite{pdg2004}; ${\bf p}_K$ is the kaon momentum in the $\phi$
meson rest frame. We then deduce $g_{\omega K^+ K^-}=g_{\phi K^+
K^-}/\sqrt{2}$ in the SU(3) symmetry limit.

The exclusive transition amplitude is
\be
\label{pp-rescat-2} {\cal M} = -\frac{i g_a g_b g_c |{\bf
p}_3|}{32 \pi^2 }\epsilon_\phi\cdot\epsilon_\omega {\cal I} ,
\ee
where the integral ${\cal I}$ has the same form as
Eq.~\ref{int-1}, but the mass $m_2$ is for the exchanged kaon.
Also, in the above equation $g_b=g_{\phi K^+ K^-}$ and
$g_c=g_{\omega K^+ K^-}$ are applied.

\section{ Intermediate vector meson rescattering}

\begin{figure}
\begin{center}
\epsfig{file=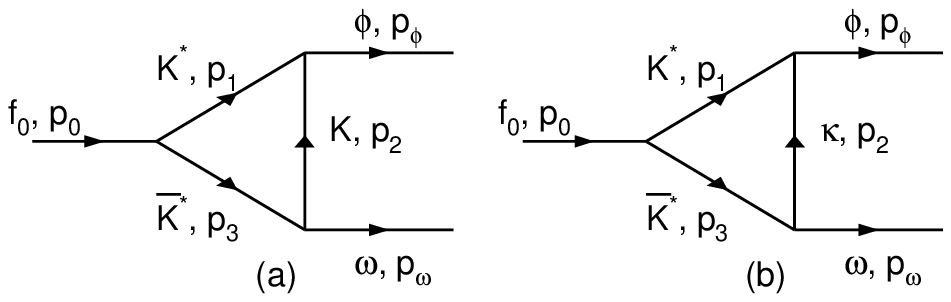, width=14cm,height=7.cm} \caption{Schematic
diagrams for intermediate meson rescatterings via (a)
$K^*\bar{K^*}$ with kaon exchanges; and (b) $K^*\bar{K^*}$ with
$\kappa$ exchange. } \protect\label{fig-2}
\end{center}
\end{figure}

In contrast with the pseudoscalar meson rescattering via vector
and pseudoscalar meson exchanges (Fig.~\ref{fig-1}(a) and (b)),
intermediate vector meson rescattering via pseudoscalar and scalar
meson exchanges, i.e. Fig.~\ref{fig-2}(a) and (b), are also
allowed.

\subsection{Rescattering via pseudoscalar meson exchange}

For transition Fig.~\ref{fig-2}(a), we consider the following
couplings:
\bea
T_a &= & -\frac{i g_a}{M_0}(p_1\cdot p_3\epsilon_1\cdot
\epsilon_3-p_1\cdot \epsilon_3 p_3\cdot\epsilon_1) \ ,\nonumber\\
T_b &\equiv & \frac{i g_b}{M_\phi}\epsilon_{\mu\nu\xi\tau}
P_\phi^\mu \epsilon_\phi^\nu p_1^\xi \epsilon_1^\tau , \nonumber\\
T_c &\equiv & \frac{i
g_c}{M_\omega}\epsilon_{\lambda\iota\kappa\sigma}P_\omega^\lambda
\epsilon_\omega^\iota p_3^\kappa \epsilon_3^\sigma ,
\eea
where $\epsilon_1$ and $\epsilon_3$ are the polarization vectors
for the rescattering vector mesons, and the coupling constants for
the $VVP$ vertices are the same as in Fig.~\ref{fig-1}(a). Similar
to the treatment for the previous two transitions, we can derive
the exclusive transition amplitude.

In the SU(3) symmetry, we can identify that the intermediate
$K^*\bar{K^*}\to \omega\phi$ via kaon exchanges could be sizeable
due to the large $\phi K^*\bar{K}$ and $\omega K^*\bar{K}$
couplings. The intermediate $\rho\rho$ scattering via pion
exchange can also couple to $\omega\phi$. Due to SU(3) symmetry
breaking, the $\phi\rho\pi$ coupling does not vanish. We adopt the
branching ratios for $\phi\to \rho\pi + 3\pi$ as an upper limit to
derive the $\phi\rho\pi$ coupling for the intermediate $\rho\rho$
scattering via pion exchanges.


The transition amplitude is
\be\label{int-vv-p}
{\cal M}^{\lambda_\phi\lambda_\omega} = \frac{i g_a g_b g_c|{\bf
p}_1|}{32\pi^2 M_\phi M_\omega M_0^2}\int
d\Omega\frac{f_{\lambda_\phi\lambda_\omega}}{p_2^2-m_2^2} {\cal
F}(p_2^2) \ ,
\ee
where $\lambda_\phi$ and $\lambda_\omega$ denote the helicity of
the final state vector meson, and function
$f_{\lambda_\phi\lambda_\omega}$ has the following expressions:
\bea
f_{00}&=& -\frac{[(P_\phi\cdot P_\omega)^2-M_0^2|{\bf
P}_\phi|^2]}{2M_\phi M_\omega} p_2^2(p_2^2-\Delta_0) \ , \nonumber\\
f_{11}&=& f_{-1-1} = \frac 12 P_\phi\cdot P_\omega
p_2^2(p_2^2-\Delta_1) \ ,
\eea
with
\bea
\Delta_0 &\equiv & \frac 12 P_\phi\cdot P_\omega
\left[3-\frac{M_\phi^2 M_\omega^2}{(P_\phi\cdot
P_\omega)^2-M_0^2|{\bf P}_\phi|^2}\right] \ , \nonumber\\
\Delta_1 &\equiv & \frac 12 P_\phi\cdot P_\omega \left[
3-\frac{M_\phi^2 M_\omega^2}{(P_\phi\cdot P_\omega)^2}\right] \ .
\eea
The integral in Eq.~(\ref{int-vv-p}) has a typical form of
\be
I=\int d\Omega \frac{p_2^2(p_2^2-\Delta)}{p_2^2-m_2^2}{\cal
F}(p_2^2) \ ,
\ee
which has been given in Ref.~\cite{zhao-eta-c}.

\subsection{Rescattering via scalar meson exchange}

The intermediate vector meson rescattering via scalar meson
exchange, Fig.~\ref{fig-2}(b), is described by the following
couplings:
\bea
T_a & =&-\frac{i g_a}{M_0}(p_1\cdot p_3\epsilon_1\cdot
\epsilon_3-p_1\cdot \epsilon_3 p_3\cdot\epsilon_1)
\nonumber\\
T_b &=& \frac{ig_b}{M_\phi}(P_\phi\cdot p_1\epsilon_\phi\cdot
\epsilon_1 -P_\phi\cdot\epsilon_1
p_1\cdot\epsilon_\phi)\nonumber\\
T_c &=& \frac{ig_c}{M_\omega} (P_\omega\cdot
p_3\epsilon_\omega\cdot\epsilon_3 -P_\omega\cdot\epsilon_3
p_3\cdot\epsilon_\omega) \ ,
\eea
where $g_a$ is the same $SVV$ coupling adopted for the $f_0 VV$
interactions in the previous subsection. $g_b$ and $g_c$ are also
$SVV$ couplings for which we apply the SU(3) flavor relation as a
constraint. For the $K^*\bar{K^*}$ rescattering, the exchanged
scalar is $\kappa$ meson. In the SU(3) symmetry, we have $g_{\phi
K^{*+}\kappa^-}=\sqrt{2}g_{\omega
K^{*+}\kappa^-}=-\sqrt{3/2}g_{\rho^0\rho^0\sigma}$ with
$g_{\rho^0\rho^0\sigma}=13.6$ broadly applied in the
literature~\cite{friman-soyeur}.

In the calculation, we assume that near the $\omega\phi$ threshold
the decay of $X$ is dominated by the $S$ wave.

\section{ Intermediate scalar meson rescattering}

\subsection{Rescattering via vector meson exchange}

\begin{figure}
\begin{center}
\epsfig{file=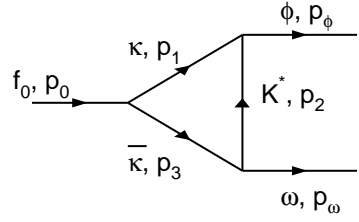, width=14cm,height=7.cm} \caption{Schematic
diagrams for intermediate $\kappa\bar{\kappa}$ rescattering  with
$K^*$ exchange. } \protect\label{fig-3}
\end{center}
\end{figure}

For the scalar meson rescattering, we consider the rescattering
via vector meson exchanges (Fig.~\ref{fig-3}). This process has
the same couplings for the $\omega$ and $\phi$ interaction
vertices as in vector meson rescattering via scalar exchanges. The
following couplings are hence applied:
\bea
T_a & =& i g_a M_0 \nonumber\\
T_b &=& \frac{i g_b}{M_\phi}(P_\phi\cdot p_2\epsilon_\phi\cdot
\epsilon_2 -P_\phi\cdot \epsilon_2 p_2\cdot\epsilon_\phi)
\nonumber\\
T_c &=& \frac{i g_c}{M_\omega} (P_\omega\cdot
p_2\epsilon_\omega\cdot\epsilon_2 -P_\omega\cdot \epsilon_2
p_2\cdot\epsilon_\omega ) \ ,
\eea
where $\epsilon_2$ is the polarization vectors of the exchanged
vector meson.

Following the standard procedure, we derive the transition
amplitude:
\be
{\cal M}^{\lambda_\phi\lambda_\omega}= \frac{i g_a g_b g_c|{\bf
p}_1|}{256\pi^2 M_\phi M_\omega M_2^2}
h_{\lambda_\phi\lambda_\omega}\int
d\Omega\frac{p_2^2(p_2^2-4m_2^2)}{p_2^2-m_2^2}{\cal F}(p_2^2) \ ,
\ee
where the function $h_{\lambda_\phi\lambda_\omega}$ has the
following form:
\bea
h_{00}&= & \frac{1}{M_\phi M_\omega}\left[(P_\phi\cdot
P_\omega)^2-M_0^2|{\bf P}_\phi|^2\right] \ , \nonumber\\
h_{11} &=& h_{-1-1} = -P_\phi\cdot P_\omega \ .
\eea

\section{ Numerical results and discussions}

To evaluate these rescatterings, we need information about the
scalar couplings to pseudoscalar, vector and scalar meson pairs,
which unfortunately are not available. If we treat the $X(1810)$
as the production of $f_0(1710)$ at higher mass tail, we then have
some additional information about $B.R.(f_0(1710)\to PP)$ from
experiment~\cite{wa102,bes-phi,bes-omega}. In particular, the
branching ratio for $f_0(1710)\to K\bar{K}$ is found to be large.
The PDG quote an average of 0.38 while the recent data from
BES~\cite{bes-phi,bes-omega} show that $B.R.(f_0(1710)\to\pi\pi
)/B.R.(f_0\to K\bar{K}))<15\%$, which lead to an estimate of
$B.R.(f_0\to K\bar{K}))\simeq 0.6$. As a result, a large coupling
for $f_0(1710)K\bar{K}$ can be expected. So far, there is no
information about the $f_0(1710)$ couplings to $VV$,
$\kappa\bar{\kappa}$, and $\sigma\sigma$. Kinematic suppressions
are also expected since the $VV$ production thresholds are
generally close to or above the $f_0(1710)$ mass. Because of
these, in our numerical analysis, we will examine the ratio of the
rescattering amplitude square over the corresponding tree
processes.

As we know that an intermediate state can generally contribute to
the transition amplitude via off-shell process. The differential
partial decay width of $J/\psi\to \gamma X\to \gamma \omega\phi$
can thus be expressed as
\bea\label{decay-1}
d\Gamma_{\omega\phi} &= & \frac{1}{(2\pi)^5}\frac{1}{ 16
M_{J/\psi}^2}\frac{1}{2J+1}\sum_{\lambda_\gamma\lambda_\phi\lambda_\omega}|{\bf
P}_\phi||{\cal M}^{\lambda_\phi\lambda_\omega}(f_0\to
\omega\phi)|^2\nonumber\\
&&\times |{\bf p}_\gamma||{\cal
M}^{\lambda_\gamma}(J/\psi\to\gamma
f_0)|^2[\mbox{Re}^2(W)+\mbox{Im}^2(W)] d W d\Omega_\gamma
d\Omega_w \ ,
\eea
where ${\bf P}_\phi$ is the momentum of $\phi$ meson in the rest
frame of $\omega$ and $\phi$; ${\bf p}_\gamma$ is the photon
momentum in the $J/\psi$ rest frame; $W$ is the invariant mass of
$\omega$ and $\phi$ system. The $f_0$ off-shell effects are taken
care by the propagator of which the real and imaginary part are:
\bea
\mbox{Re}(W)&=& \frac{W^2-M_0^2}{(W^2-M_0^2)^2+M_0^2\Gamma_T^2}
\nonumber\\
\mbox{Im}(W)&=& -\frac{M_0\Gamma_T}{(W^2-M_0^2)^2+M_0^2\Gamma_T^2}
\,
\eea
where $M_0$ and $\Gamma_T$ are the mass and total width of the
$f_0$ state.

Similarly, we can express the differential partial decay width for
$J/\psi\to \gamma f_0\to \gamma K\bar{K}$ as
\bea
d\Gamma_{K\bar{K}} &= & \frac{1}{(2\pi)^5}\frac{1}{ 16
M_{J/\psi}^2}\frac{1}{2J+1}\sum_{\lambda_\gamma}|{\bf P}_k||{\cal
M}(f_0\to K\bar{K})|^2\nonumber\\
&&\times |{\bf p}_\gamma||{\cal
M}^{\lambda_\gamma}(J/\psi\to\gamma
f_0)|^2[\mbox{Re}^2(W)+\mbox{Im}^2(W)] d W d\Omega_\gamma
d\Omega_w \ ,
\eea

Note that to produce the off-shell $f_0$ at mass $W$ means that
the invariant transition matrix element ${\cal
M}^{\lambda_\gamma}(J/\psi\to\gamma f_0)$ is $W$-dependent. To
obtain the partial decay width we also need to know the
information about the $J/\psi\to \gamma f_0$ transition. There are
phenomenological studies available in the literature for
$J/\psi\to \gamma f_0(1710)$~\cite{cfl}. Here, we are interested
in the decays of $f_0\to\omega\phi$ via different rescattering
processes. To eliminate the ambiguities from the first vertex, we
take the ratio between these two differential partial decay widths
for those above-listed meson rescatterings, and define, for
example, for the pseudoscalar meson rescattering via vector meson
exchange:
\be
R_V^{PP}\equiv \frac{|{\bf P}_\phi|
\sum_{\lambda_\phi\lambda_\omega}|{\cal
M}_V^{\lambda_\phi\lambda_\omega}(f_0\to PP\to\omega\phi)
|^2}{|{\bf P}_k||{\cal M}(f_0\to PP)|^2} \ ,
\ee
where the superscript $PP$ denotes the rescattered pseudoscalar
meson pair and the subscript $V$ denotes the exchanged vector
meson; ${\bf P}_k$ is the three momentum carried by the
pseudoscalar meson in the c.m. frame of $f_0$ with mass $W$. By
scanning over a range of $W$, the energy dependence of the ratio
will reveal the evolution of the intermediate meson loop.

Similarly, we define the ratios for other rescatterings relative
to the corresponding tree processes:
\be
R_P^{PP}\equiv \frac{|{\bf P}_\phi|
\sum_{\lambda_\phi\lambda_\omega}|{\cal
M}_P^{\lambda_\phi\lambda_\omega}(f_0\to PP\to\omega\phi)
|^2}{|{\bf P}_k||{\cal M}(f_0\to PP)|^2} \
\ee
for the pseudoscalar meson rescattering via pseudoscalar meson
exchange;
\be
R_P^{VV}\equiv \frac{|{\bf P}_\phi|
\sum_{\lambda_\phi\lambda_\omega}|{\cal
M}_P^{\lambda_\phi\lambda_\omega}(f_0\to VV\to\omega\phi)
|^2}{|{\bf P}_v||{\cal M}(f_0\to VV)|^2} \
\ee
for vector meson rescattering via pseudoscalar meson exchange;
\be
R_S^{VV}\equiv \frac{|{\bf P}_\phi|
\sum_{\lambda_\phi\lambda_\omega}|{\cal
M}_S^{\lambda_\phi\lambda_\omega}(f_0\to VV\to\omega\phi)
|^2}{|{\bf P}_v||{\cal M}(f_0\to VV)|^2} \
\ee
for vector meson rescattering via scalar meson exchange; and
\be
R_S^{SS}\equiv \frac{|{\bf P}_\phi|
\sum_{\lambda_\phi\lambda_\omega}|{\cal
M}_S^{\lambda_\phi\lambda_\omega}(f_0\to SS\to\omega\phi)
|^2}{|{\bf P}_s||{\cal M}(f_0\to SS)|^2} \
\ee
for scalar meson rescattering via scalar meson exchange.

It is also interesting to examine the evolution of the meson loop
in terms of $W$ compared with the tree processes for $f_0\to
K\bar{K}$. Therefore, we define ratio $Q$ as follows:
\bea
Q_P^{VV}&\equiv &\frac{|{\bf P}_\phi|
\sum_{\lambda_\phi\lambda_\omega}|{\cal
M}_P^{\lambda_\phi\lambda_\omega}(f_0\to VV\to\omega\phi)
|^2}{|{\bf P}_k||{\cal M}(f_0\to PP)|^2} \ , \\
Q_S^{VV}&\equiv &\frac{|{\bf P}_\phi|
\sum_{\lambda_\phi\lambda_\omega}|{\cal
M}_S^{\lambda_\phi\lambda_\omega}(f_0\to VV\to\omega\phi)
|^2}{|{\bf P}_k||{\cal M}(f_0\to PP)|^2} \ , \\
Q_S^{SS}&\equiv &\frac{|{\bf P}_\phi|
\sum_{\lambda_\phi\lambda_\omega}|{\cal
M}_S^{\lambda_\phi\lambda_\omega}(f_0\to SS\to\omega\phi)
|^2}{|{\bf P}_k||{\cal M}(f_0\to PP)|^2} \ ,
\eea
where ${\bf P}_k$ is the three momenta for $f_0\to PP$ in the
$f_0$ rest frame with mass $W$. Due to lack of information about
the $f_0 VV$ and $f_0 SS$ couplings, the ratio $Q$ will possess
large uncertainties.

\begin{figure}
\begin{center}
\epsfig{file=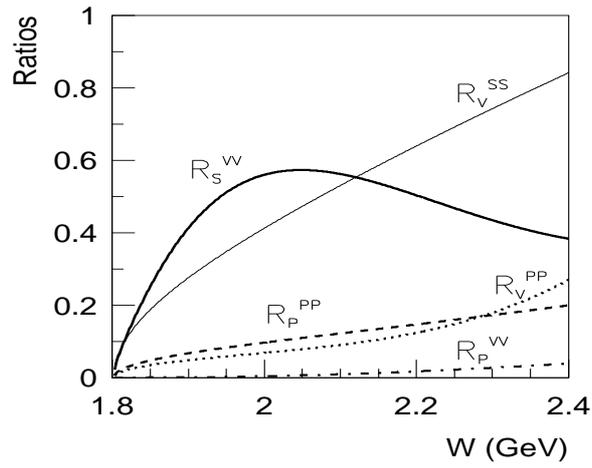, width=10cm,height=8.cm} \caption{The
evolution of the ratios of the invariant amplitude square of the
meson loops to the corresponding tree processes in terms of the
initial scalar meson masses. Definition of the ratios is given in
the text. } \protect\label{fig-4}
\end{center}
\end{figure}

The numerical results for ratio $R$ are presented in
Fig.~\ref{fig-4}. The $K^*\bar{K^*}$ rescattering via scalar meson
exchange turns out to be the largest contribution as denoted by
the thick solid line. A strong enhancement is produced with an
increasing invariant mass for the $\omega\phi$ system. Since this
is the ratio between the $K^*\bar{K^*}\kappa$ loop and the tree
process for $f_0\to K^*\bar{K^*}$, it is independent of the $f_0
K^*\bar{K^*}$ coupling. As shown by the solid curve, the
$K^*\bar{K^*}$ rescattering via $\kappa$ exchange has a sizeable
fraction compared with the $f_0\to K^*\bar{K^*}$ transition.

To estimate this channel's contributions to the $\omega\phi$
partial width, we assume that ${\cal
M}^{\lambda_\gamma}(J/\psi\to\gamma f_0)$ is insensitive to $W$.
Eq.~(\ref{decay-1}) can then be expressed as
\be
\Gamma_{\omega\phi}=\Gamma_{J/\psi\to\gamma f_0}\times
B.R.(f_0\to\omega\phi) \ ,
\ee
where
\be
B.R.(f_0\to\omega\phi)\equiv \int\int\frac{|{\bf
P}_\phi|}{16\pi^3} \sum_{\lambda_\phi\lambda_\omega}|{\cal
M}^{\lambda_\phi\lambda_\omega}(f_0\to \omega\phi)|^2
[\mbox{Re}^2(W)+\mbox{Im}^2(W)] d W  d\Omega_w \ .
\ee

For $K^*\bar{K^*}$ rescattering via scalar meson exchange, we
examine the following conditions:

(i) With $M_\kappa=0.7$ GeV and $B.R.(f_0\to K^*\bar{K^*})=0.1$,
we find $\Gamma_{\omega\phi}=1.36\sim 3.01$ MeV for a mass range
of $f_0$ from $1.74\sim 1.81$ GeV.

(ii) With $M_\kappa=0.7$ GeV and fixing the mass of the $f_0$ at
1.74 GeV, for a range of $B.R.(f_0\to K^*\bar{K^*})=0.1\sim 0.3$,
we obtain $\Gamma_{\omega\phi}=1.36\sim 4.09$ MeV, which
correspond to $B.R.(f_0\to\omega\phi)=(0.97\sim 2.92)\%$. For the
PDG value $B.R.(J/\psi\to \gamma f_0(1710)\to \gamma
K\bar{K})=8.5\times 10^{-4}$ and BES estimate of
$B.R.(f_0(1710)\to K\bar{K})=0.6$, we derive $B.R.(J/\psi\to
\gamma f_0(1710))=1.4\times 10^{-3}$. Thus, we estimate
$B.R.(J/\psi\to\gamma f_0\to \gamma \omega\phi)\simeq (1.36\sim
4.09)\times 10^{-5}$.

(iii) By fixing the mass of the $f_0$ at 1.74 GeV, and
$B.R.(f_0\to K^*\bar{K^*})=0.1$, and then varying the mass of the
$\kappa$ from 0.8 - 0.6 GeV, we obtain
$\Gamma_{\omega\phi}=0.45\sim 3.61$ MeV corresponding to
$B.R.(J/\psi\to\gamma f_0\to \gamma \omega\phi)\simeq (0.45\sim
3.61)\times 10^{-5}$. The calculation results turn to be sensitive
to the mass of the exchanged $\kappa$ meson, which still has large
uncertainties~\cite{pdg2004}. We also find that
$\Gamma_{\omega\phi}$ drops fast with an increasing $\kappa$ mass.
With $M_\kappa >0.9$ GeV, the branching ratio for $f_0\to
\omega\phi$ will be just about 0.01\%. On the other hand, although
a smaller mass for $\kappa$ can produce relatively large branching
ratios for $\omega\phi$ channel, $M_\kappa=0.6$ GeV can be
regarded as the lower bound for the $\kappa$ mass, which gives
$B.R.(f_0\to \omega\phi)\simeq 2.6\%$.

In Fig.~\ref{fig-4} the dashed line is the ratio $R_P^{PP}$ for
pseudoscalar meson rescattering via pseudoscalar meson exchange,
and the dotted and dot-dashed line are for $R_V^{PP}$ and
$R_P^{VV}$, respectively. These ratios are found flat in terms of
the increasing invariant mass. In particular, $R_P^{VV}$ is found
negligible at low $W$. After considering the Breit-Wigner factor,
their contributions to the $\omega\phi$ width are negligible.

The thin solid line denotes $R_V^{SS}$ for the scalar meson
rescattering via vector meson exchange. It shows a rapid increase
along the invariant mass $W$. We also derive its contributions to
$f_0\to\omega\phi$ partial width, and find
$B.R.(f_0\to\omega\phi)< 1\%$. This suggests that the Breit-Wigner
factor will kill the invariant amplitudes quickly when the scalar
goes off-shell. It should be noted that so far there are no data
available for $f_0(1710)\to \kappa\bar{\kappa}$ in experiment. We
have assumed $B.R.(f_0\to \kappa\bar{\kappa})=0.1\sim 0.2$ in the
calculation. However, this does not bring significant
contributions to the $\omega\phi$ final state. Although the
$\kappa\bar{\kappa}$ rescattering is unlikely to have large
contributions to $\omega\phi$, it may becomes important in heavier
scalar decays. The recent BES analysis shows $\kappa$ signals in
$J/\psi\to \bar{K^*}\kappa$~\cite{bes-kstarkappa,lihb}, which
suggests strong $\kappa$ couplings in some channels. As shown in
Fig.~\ref{fig-4}, its increasing contributions in the
rescatterings imply that it may play a role in OZI violation
processes at higher energies.

\begin{figure}
\begin{center}
\epsfig{file=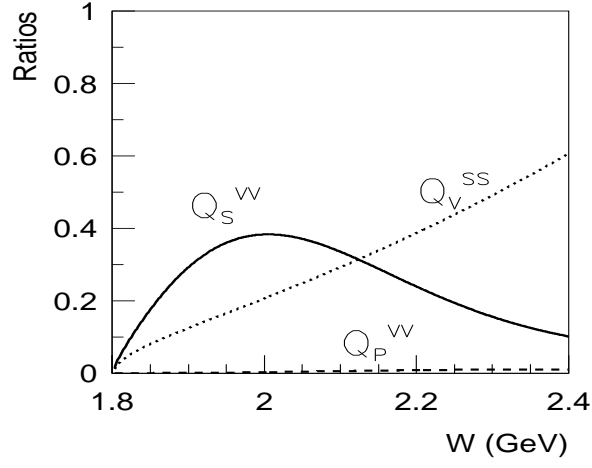, width=10cm,height=8.cm} \caption{The
evolution of the ratios of the invariant amplitude square of the
meson loops to $f_0\to K\bar{K}$ in terms of the initial scalar
meson masses. Definition of the ratios is given in the text.}
\protect\label{fig-5}
\end{center}
\end{figure}

The ratio $Q$ for the intermediate pseudoscalar meson
rescatterings are the same as $R$. So we only present the ratios
$Q_S^{VV}$, $Q_P^{VV}$ and $Q_V^{SS}$ in Fig.~\ref{fig-5}. It
shows that the $K^*\bar{K^*}$ rescattering via $\kappa$ exchange
is the major contribution to the $\omega\phi$ partial width since
it has a steep increase at low $W$ where the off-shell effects are
relatively small. The $\kappa\kappa$ rescattering increases with
the increasing $W$ as shown by the dotted curve. However, due to
the suppression from the off-shell factors, it does not
significantly contribute to the $f_0\to\omega\phi$ partial width.
Ratio $Q_P^{VV}$ illustrated by the dashed curve also turns out to
be negligible.

\begin{figure}
\begin{center}
\epsfig{file=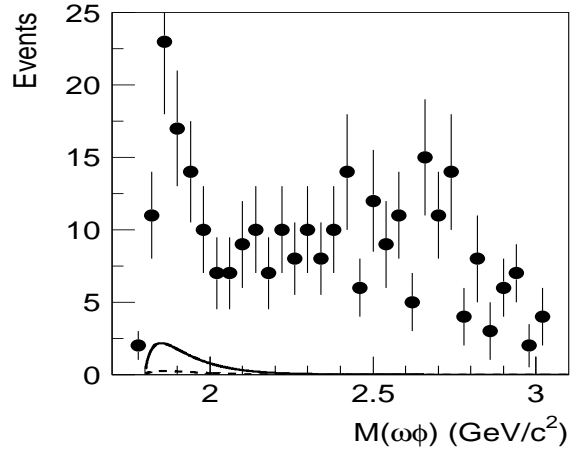, width=10cm,height=8.cm} \caption{Invariant
mass spectrum for $\omega\phi$ in comparison with the BES
experimental data~\cite{bes-2006}. The solid curve denotes the
$\omega\phi$ invariant mass distribution with $B.R.(f_0\to
K^*\bar{K^*})=0.3$ and $M_\kappa=0.6$ GeV, while the dashed curve
with $B.R.(f_0\to K^*\bar{K^*})=0.1$ and $M_\kappa=0.7$ GeV. }
\protect\label{fig-6}
\end{center}
\end{figure}

We also study the $\omega\phi$ invariant mass spectrum to examine
the behavior of the intermediate meson rescatterings and compare
it with the BES data~\cite{bes-2006}. In Fig.~\ref{fig-6}, the
solid curve denotes the $\omega\phi$ invariant mass distribution
with $B.R.(f_0\to K^*\bar{K^*})=0.3$ and $M_\kappa=0.6$ GeV, which
is the higher bound for $K^*\bar{K^*}$ rescattering via $\kappa$
exchange. The dashed curve denotes the results with $B.R.(f_0\to
K^*\bar{K^*})=0.1$ and $M_\kappa=0.7$ GeV.
The peak of the enhancement sits around 1.85 GeV which seems to be
consistent with the data. Note that though the partial width,
$\Gamma(f_0\to\omega\phi)=10.8$ MeV (derived with $B.R.(f_0\to
K^*\bar{K^*})=0.3$ and $M_\kappa=0.6$ GeV) is still significantly
smaller than the Breit-Wigner fit $\Gamma(X(1810)\to
\omega\phi)=105\pm 20$ MeV, it gives $B.R.(J/\psi\to\gamma f_0\to
\gamma \omega\phi)\simeq 1.08\times 10^{-4}$, which is comparable
with the experimental data~\cite{bes-2006}. This shows that if the
enhancement does originate from the $f_0(1710)$ due to
intermediate meson rescatterings, it will exhibit a rather narrow
width. In this sense, a broad width at order of 100 MeV may favor
its being a real Breit-Wigner resonance.

\section{Summary}

In summary, based on the present experimental information we
examine the intermediate meson rescattering contributions to
$J/\psi\to\gamma X\to \gamma \omega\phi$ by assuming that
$X=f_0(1710)$ with a mass at 1.74$\sim$ 1.81 GeV. We find that the
contributions from the vector meson $K^*\bar{K^*}$ rescattering
via scalar meson exchange can produce some enhancement near the
$\omega\phi$ threshold. The other intermediate meson
rescatterings, such as pseudoscalar and scalar meson
rescatterings, are all found relatively small.

The calculation results for $B.R.(J/\psi\to\gamma f_0\to \gamma
\omega\phi)$ range from $(1.36\sim 10.8)\times 10^{-5}$ for
different values of $B.R.(f_0\to K^*\bar{K^*})$ and the $\kappa$
mass. The derived partial decay width for $f_0\to \omega\phi$
turns out to be at least one order-of-magnitude smaller than the
observed partial width for $X(1810)\to
\omega\phi$~\cite{bes-2006}. This seems to make it unlikely that
the observed enhancement in $\omega\phi$ invariant mass spectrum
is from intermediate meson rescatterings in $f_0(1710)$ decays.
However, due to lack of information about the $f_0(1710)\to SS$
and $VV$, we find it does not suffice to conclude on the nature of
the enhancement. There still exist uncertainties in our
calculations which should be cautioned: (i) the value of
$\Gamma_{\omega\phi}$ turns to be sensitive to the choice of
cut-off energies for the dipole form factor, and it can lead to a
change about a factor of 2; (ii) the contributions from the
intermediate meson rescatterings may be underestimated due to the
application of the on-shell approximation. Under such an
approximation, only the imaginary part of the transition
amplitudes is picked up. For the $VV$ rescattering, since the
vector meson pair is close to the $f_0$ threshold, the on-shell
approximation will introduce double kinematic suppressions to the
imaginary part, while the real part will not suffer such a
suppression. Due to this, the inclusion of the real part could
enhance the partial width to $\omega\phi$ via $K^*\bar{K^*}$
rescatterings. This issue should be studied in the future with
more knowledge about the $VVS$ couplings available.

It is necessary to have more experimental information about the
signals of $X(1810)$ in $PP$ and $VV$ (i.e. $K^*\bar{K^*}$,
$\omega\omega$ and $\rho\rho$). In particular, a direct analysis
of $J/\psi\to\phi X\to \phi\omega\phi$ and $J/\psi\to \omega
X\to\omega\omega\phi$ could be useful for establishing the
$X(1810)$ as a real resonance. In case that $X(1810)$ is a new
scalar, it will be a challenge for theory in the understanding of
the scalar meson spectrum, and may also be a chance for us to gain
more insights into the underlying dynamics.

\section*{Acknowledgement}

The authors thank F.E. Close for useful discussions. This work is
supported, in part, by the U.K. Engineering and Physical Sciences
Research Council (Grant No. GR/S99433/01), and the National Nature
Science Foundation of China (Grant No. 10225525 and 10435080).


\end{document}